\documentclass[aps,prl,showpacs,twocolumn]{revtex4}
\usepackage{amssymb}

\usepackage{graphicx}

\input{tcilatex}

\begin{document}

\title{Disentangling topological degeneracy in the entanglement spectrum of
one-dimensional symmetry protected topological phases}
\author{Wen-Jia Rao$^{1}$, Guang-Ming Zhang$^{1,2}$, and Kun Yang$^{3}$}
\affiliation{$^{1}$State Key Laboratory of Low-Dimensional Quantum Physics and Department
of Physics, Tsinghua University, Beijing 100084, China.\\
$^{2}$Collaborative Innovation Center of Quantum Matter, Beijing, China\\
$^{3}$National High Magnetic Field Laboratory and Physics Department,
Florida State University, Tallahassee, Florida 32310, USA}
\date{\today }

\begin{abstract}
One-dimensional valence bond solid (VBS) states represent the simplest
symmetry protected topological phases. We show that their ground state
entanglement spectrum contains both topological and non-topological
structures. For the $SO(3)$ symmetric VBS states with odd-integer spins, the
two-fold topological degeneracy is associated with an underlying $%
Z_{2}\times Z_{2}$ symmetry that protects the corresponding topological
phase. In general, for the $SO(2S+1)$ symmetric VBS states with integer
spins $S$, the corresponding protecting symmetry is identified as the $%
(Z_{2}\times Z_{2})^{S}$ symmetry, yielding the $2^{S}$-fold topological
degeneracy. The topological degeneracy and associated protecting symmetry
can be identified by a non-local unitary transformation, which changes the
topological order of the VBS states into conventional ferromagnetic order.
\end{abstract}

\maketitle

\section{Introduction}

Topological properties of low-dimensional quantum many-body systems have
been attracting considerable interest in both quantum information sciences
and condensed matter physics. Remarkably, it is understood that important
information of the topological phase is encoded in the von Neumann
entanglement entropy of its ground state.\cite%
{Kitaev-Preskill,Levin-Wen,Zhang-Vishwanath} In a seminal paper, Li and
Haldane\cite{Li-Haldane} proposed that the entanglement spectrum (ES)
labeled by the quantum numbers of symmetries preserved by the Schmidt
singular-value decomposition contains \emph{more} information than the
entanglement entropy. In particular, they found the largest eigenvalues of
the ground state reduced density matrix mimics the physics edge spectrum of
a gapped quantum system, thus revealing the bulk topological order. A large
number of works have followed, suggesting that the ''low-energy''structure
of ES reveals the bulk topological order.\cite%
{Fidkowski,cirac-2011,Bernevig-2011,Qi-2012}

Recently, a critique on this body of work was given by Chandran, Khemani,
and Sondhi\cite{Chandran-Sondhi}, who posed a question,``How universal is
the entanglement spectrum?'' In this paper we will address this important
question by studying the ES of a family of one-dimensional symmetry
protected topological (SPT) phases.\cite%
{Chen-Gu-Wen-2011,Schuch,chen-gu-liu-wen} The SPT phases have robust gapless
edge excitations and can not be continuously connected to trivial product
state without either breaking the protecting symmetry or closing the energy
gap. Moreover, the one-dimensional SPT phases have been generally classified
by group cohomologies completely.\cite{Chen-Gu-Wen-2011,Schuch,Quella} ES of
these phases are sufficiently simple and can be calculated analytically for
model wave functions\cite{Katsura-2007}, thus providing an ideal setting for
addressing this issue. Our results demonstrate that ES of topological phases
include both topological/universal and non-topological parts. In certain
cases the universal part can be isolated, and is extremely useful in
revealing the bulk topological order.

The simplest example of SPT phases is the Haldane gapped phase of the
antiferromagnetic spin-1 chain,\cite{Haldane-1983} which is protected by any
one of the following discrete symmetries: time reversal symmetry, link
inversion symmetry, or the $D_{2}\backsimeq Z_{2}\times Z_{2}$ symmetry
comprising $\pi $ rotations about two orthogonal axes.\cite%
{Gu-Wen-2009,Pollmann-Oshikawa-2010,Pollmann-Oshikawa-2012} The fixed point
wave functions for these phases are the valence bond solid (VBS) states.\cite%
{AKLT} We show their ES contains a topological part, which is universal, and
a non-topological or non-universal part. The topological part can be
isolated by a topological disentangler, a non-local unitary transformation.
For the familiar $SO(3)$ symmetric VBS states with odd-integer spins, their
ES is given by a single energy level with $\left( S+1\right) $-fold
degeneracy. However, the bulk topological order only gives rise to (or
protects) a two-fold degeneracy, which is associated with the $Z_{2}\times
Z_{2}$ symmetry. The remaining $(S+1)/2$-fold degeneracy is \emph{not}
protected by the bulk topological order.

Furthermore, for VBS phases with higher symmetry, we can find corresponding
high topological degeneracy and associated protecting symmetry via a
generalized non-local unitary transformation. Specifically, we find in the $%
SO(5)$ symmetric VBS state with spin-$2$,\cite{Tu-2008,TZX} the topological
degeneracy are four-fold and the protecting symmetry can be identified as $%
(Z_{2}\times Z_{2})^{2}$. In general, for the $SO(2S+1)$ symmetric VBS
states with integer spin-$S$,\cite%
{Tu-2008,TZX,Tu-2009,Else-Bartlett-Doherty,Quella-2013} the corresponding
protecting symmetry is given by $(Z_{2}\times Z_{2})^{S}$, and the
topological degeneracy is $2^{S}$-fold.

\section{VBS states and entanglement spectrum\textit{\ }}

The $SO(3)$ symmetric VBS states with arbitrary integer spin for a periodic
spin chain is given by\cite{AKLT} 
\begin{equation}
{\small \left| \text{VBS}\right\rangle =\prod_{i=1}^{N}\left( a_{i}^{\dagger
}b_{i+1}^{\dagger }-b_{i}^{\dagger }a_{i+1}^{\dagger }\right) ^{S}\left| 
\text{vac}\right\rangle ,}
\end{equation}%
where the Schwinger spin-boson representation with a local constraint $%
a_{i}^{\dagger }a_{i}+b_{i}^{\dagger }b_{i}=2S$ has been used as%
\[
{\small s_{i}^{+}=a_{i}^{\dagger }b_{i},s_{i}^{-}=b_{i}^{\dagger
}a_{i},s_{i}^{z}=\frac{1}{2}\left( a_{i}^{\dagger }a_{i}-b_{i}^{\dagger
}b_{i}\right) .} 
\]%
One can construct an $SU(2)$ spin invariant model Hamiltonians, whose ground
states are exactly given by these VBS states.\cite{Arovas-Haldane-1988} In
the thermodynamic limit, the spin correlation function decays exponentially
with a correlation length $\xi =1/\ln \left( 1+2/S\right) $, implying that
the VBS state is a gapped disordered state.

For a finite length chain with open boundary condition, the ground state
wave function is expressed as{\small 
\begin{eqnarray}
&&|\text{VBS}\left( \tau _{1}^{z}=\alpha ,\tau _{N}^{z}=\beta \right)
\rangle =\left( a_{1}^{\dagger }\right) ^{\frac{S}{2}+\alpha }\left(
b_{1}^{\dagger }\right) ^{\frac{S}{2}-\alpha }  \nonumber \\
&&\times \prod_{i=1}^{N-1}\left( a_{i}^{\dagger }b_{i+1}^{\dagger
}-b_{i}^{\dagger }a_{i+1}^{\dagger }\right) ^{S}\left( a_{N}^{\dagger
}\right) ^{\frac{S}{2}+\beta }\left( b_{N}^{\dagger }\right) ^{\frac{S}{2}%
-\beta }|\text{vac}\rangle .
\end{eqnarray}%
}where $\alpha $ and $\beta $ can be chosen from $-S/2$,$-S/2+1, \cdots, S/2$%
, representing the eigenvalues of the edge spins of $\tau _{1}^{z}$ and $%
\tau _{N}^{z}$. There are total $\left( S+1\right) ^{2}$ degenerate edge
states.

As model wave functions of a family of one-dimensional SPT phases, the VBS
states exhibit the simplest ES. To calculate the ES for the wave functions $%
| $VBS$\left( \alpha ,\beta \right) \rangle $, we cut the spin chain into
two parts A and B (see Fig.1a), and the VBS wave function can be expressed
as {\small 
\begin{eqnarray}
&&|\text{VBS}\left( \alpha ,\beta \right) \rangle  \nonumber \\
&=&\sum_{n=0}^{S}\left( -1\right) ^{S-n}C_{S}^{n}\left| \text{VBS}\left(
\tau _{1}^{z}=\alpha ,\tau _{L}^{z}=n-S/2\right) \right\rangle _{A} 
\nonumber \\
&&\text{ \ }\times \left| \text{VBS}\left( \tau _{L+1}^{z}=S/2-n,\tau
_{N}^{z}=\beta \right) \right\rangle _{B},
\end{eqnarray}%
}where $C_{S}^{n}=S!/n!(S-n)!$. After the normalization of all wave
functions are carefully considered, the resulting wave function is written
as {\small 
\begin{eqnarray}
&&|\text{VBS}\left( \alpha ,\beta \right) \rangle =\sum_{n=0}^{S}\left(
-1\right) ^{S-n}e^{-\xi /2}  \nonumber \\
&&\text{ \ \ \ \ \ }\times \left| \text{VBS}\left( \alpha ,n-S/2\right)
\right\rangle _{A}\left| \text{VBS}\left( S/2-n,\beta \right) \right\rangle
_{B},
\end{eqnarray}%
}where a ratio of the length dependent factors has been used: $\frac{C(N)}{%
C(N_{A})C(N_{B})}=\left( S!\right) ^{2}(S+1)$, and the entanglement energy
level is just given by a single level $\zeta =\ln (S+1)$ with a degeneracy $%
(S+1)$, which exactly corresponds to the degeneracy of the VBS wave function
with one edge spin. So the degeneracy of each edge spin and the degeneracy
of the entanglement level has one-to-one correspondence.\cite{Li-Haldane}
However, a natural question to ask is whether all the degeneracies of the
entanglement energy level are topologically protected or not.

\begin{figure}[t]
\includegraphics[width=7.5cm]{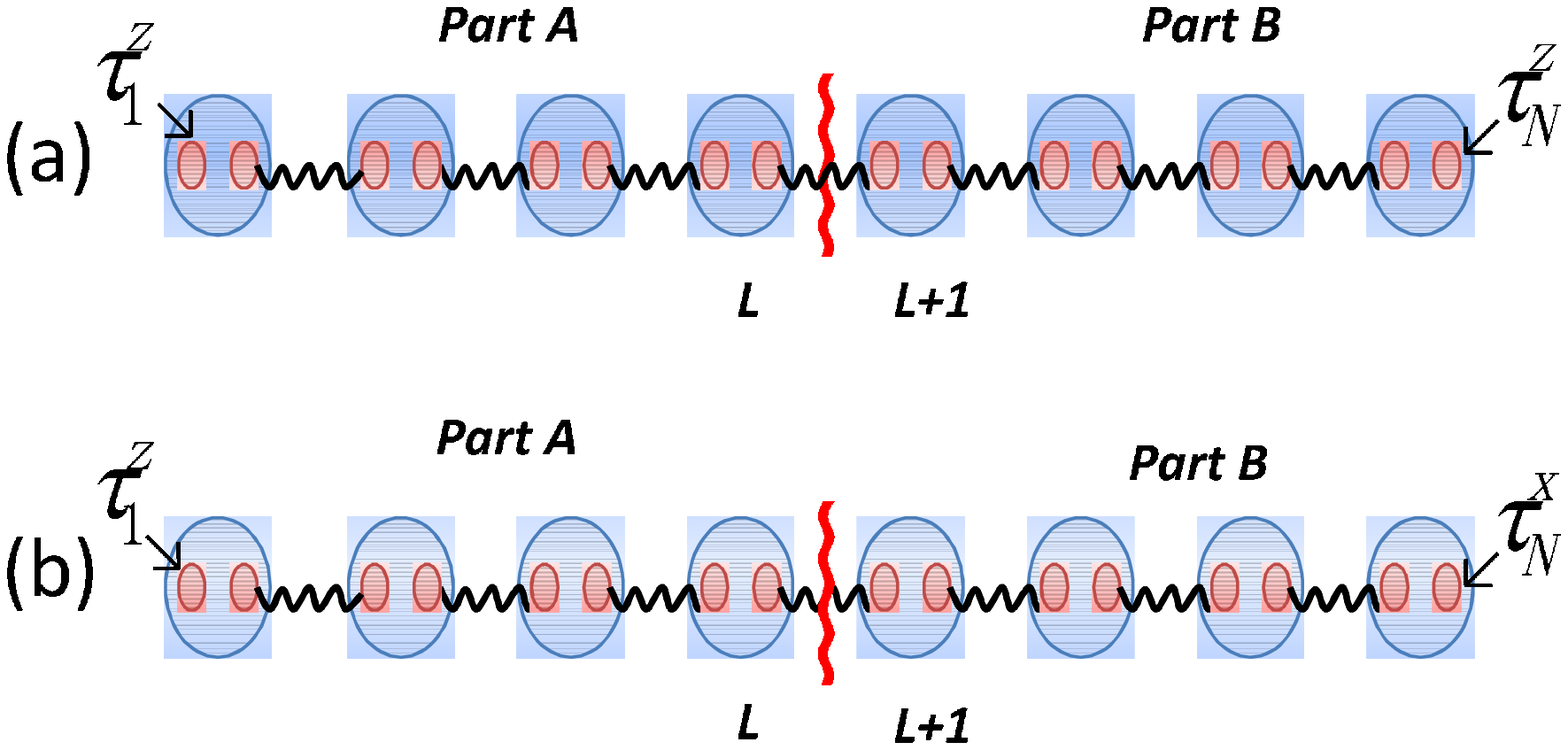}
\caption{(Color online) The VBS states with open boundary conditions in the
different edge spin representations. (a) The right edge spin is specified by
its $\protect\tau_z$ eigenvalue; (b) The right edge spin is specified by its 
$\protect\tau_x$ eigenvalue.}
\end{figure}

\section{Non-local unitary transformation}

In order to characterize the topological features of these VBS states,
non-local string order parameters (SOP) were first proposed for the spin-1
case,\cite{Nijs-Rommelse}, and then generalized to arbitrary integer spins%
\cite{Oshikawa-1992,Totsuka-Suzuki}: 
\begin{equation}
{\small O_{s}^{\mu }\left( j,k\right) =\lim_{\left| k-j\right| \rightarrow
\infty }\langle }\text{{\small VBS}}{\small {|}}s_{j}^{\mu }\exp \left( 
{\small i\pi \sum_{i=j}^{k-1}s_{i}^{\mu }}\right) {\small s_{k}^{\mu }|}%
\text{{\small VBS}}{\small \rangle }
\end{equation}%
where $\mu =z$, $x$ or $y$. In the thermodynamic limit, the SOP is found to
be 
\begin{equation}
{\small O_{s}^{\mu }\left( j,k\right) =\left( \frac{S+1}{S+2}\right)
^{2}\delta _{S,\text{odd}},}
\end{equation}%
which reveals the fundamental difference between VBS states with odd- and
even-integer spins. Only VBS states with odd-integer spins represent
topologically nontrivial phases. An alternative and more useful
characterization was revealed through a non-local unitary transformation\cite%
{Oshikawa-1992,Kennedy-Tasaki-1992} 
\begin{equation}
{\small U=\prod_{k=2}^{N}\prod_{j=1}^{k-1}e^{i\pi s_{j}^{z}s_{k}^{x}},}
\end{equation}%
from which the local spin operators are transformed to {\small 
\begin{eqnarray}
\widetilde{s}_{j}^{z} &=&U^{-1}s_{j}^{z}U=s_{j}^{z}\text{exp}\left( i\pi
\sum_{i=1}^{j-1}s_{i}^{z}\right) ,  \nonumber \\
\widetilde{s}_{j}^{x} &=&U^{-1}s_{j}^{x}U=s_{j}^{x}\text{exp}\left( i\pi
\sum_{i=j+1}^{N}s_{i}^{x}\right) ,
\end{eqnarray}%
}where a string operator in $\widetilde{s}_{j}^{z}$ involves all the sites
from left edge site up to $j-1$, while the string operator in $\widetilde{s}%
_{j}^{x}$ involves all the sites from the site $j+1$ up to the right edge
site. These properties will become crucial in studying the \emph{transformed}
VBS wave function. Meanwhile, the non-local SOP becomes to the local spin
correlation via this transformation: 
\begin{equation}
{\small U^{-1}s_{j}^{\mu }s_{k}^{\mu }U=s_{j}^{\mu }\exp \left( i\pi
\sum_{i=j}^{k-1}s_{i}^{\mu }\right) s_{k}^{\mu },\text{ \ \ }\mu =z,x.}
\end{equation}

\section{Transformed VBS states and its ES}

For the convenience of performing the non-local unitary transformation, the
VBS wave functions are expressed in terms of the eigenvalue basis of the
left edge spin $\tau _{1}^{z}$ and the right edge spin in the $\tau _{N}^{x}$%
: {\small 
\begin{eqnarray*}
&&|\text{VBS}\left( \tau _{1}^{z}=\alpha ,\tau _{N}^{x}=\beta \right)
\rangle =\left( a_{1}^{\dagger }\right) ^{\frac{S}{2}+\alpha }\left(
b_{1}^{\dagger }\right) ^{\frac{S}{2}-\alpha } \\
&&\times \prod_{i=1}^{N-1}\left( a_{i}^{\dagger }b_{i+1}^{\dagger
}-b_{i}^{\dagger }a_{i+1}^{\dagger }\right) ^{S}\left( c_{N}^{\dagger
}\right) ^{\frac{S}{2}+\beta }\left( d_{N}^{\dagger }\right) ^{\frac{S}{2}%
-\beta }|\text{vac}\rangle ,
\end{eqnarray*}%
}where $c_{j}^{\dagger }=\left( a_{j}^{\dagger }+b_{j}^{\dagger }\right) /%
\sqrt{2}$ and $d_{j}^{\dagger }=\left( a_{j}^{\dagger }-b_{j}^{\dagger
}\right) /\sqrt{2}$. The transformed VBS states are ferromagnetic long-range
ordered states with two finite magnetizations $\widetilde{m}_{j}^{x}$ and $%
\widetilde{m}_{j}^{z}$ as local order parameters\cite{Rao-Zhang}:%
\begin{eqnarray*}
{\small \langle \widetilde{VBS}\left( \alpha ,\beta \right) |s_{j}^{z}|%
\widetilde{VBS}\left( \alpha ,\beta \right) \rangle }{\small =\left(
-1\right) ^{\alpha +S/2}\left( \frac{S+1}{S+2}\right) \delta _{S.\text{odd}},%
} && \\
{\small \langle \widetilde{VBS}\left( \alpha ,\beta \right) |s_{j}^{x}|%
\widetilde{VBS}\left( \alpha ,\beta \right) \rangle }{\small =\left(
-1\right) ^{\beta +S/2}\left( \frac{S+1}{S+2}\right) \delta _{S.\text{odd}}.}
&&
\end{eqnarray*}%
Therefore the non-local transformation with the discrete symmetry $%
Z_{2}\times Z_{2}$ plays the role of a topological disentangler, because it
turns topological order of the original VBS state into conventional
ferromagnetic order. The eigenvalues $\alpha $ and $\beta $ of the edge
spins are used to classify the transformed degenerate ground states $|%
\widetilde{\text{VBS}}\left( \alpha ,\beta \right) \rangle $. There exist
four different classes of long-range ordered ferromagnetic states with the
degeneracy $\left( S+1\right) /2$ each. These degeneracies of each
ferromagnetic ordered state only appear for $S>1$ odd-integer spins and
results from the partial polarization of the spins. The detailed
ferromagnetic configurations are summarized in Table I. 
\begin{table}[t]
\caption{The transformed SO(3) symmetric VBS states with odd-integer spins.
They are ferromagnetic states classified into four groups by the eigenvalues
of the edge spins $\protect\tau _{1}^{z}$ and $\protect\tau _{N}^{x}$, each
with $(S+1)/2$-fold degeneracy.}{\small 
\begin{ruledtabular}
\begin{tabular}{cccc}
left edge $\tau_1^{z}=\alpha$ & right edge $\tau_N^x=\beta $ &  $\tilde{m}_j^z$ & $\tilde{m}_j^x$ \\
\hline $\frac{S}{2},\frac{S}{2}-2,...,-\frac{S}{2}+1$ &
$\frac{S}{2},\frac{S}{2}-2,...,-\frac{S}{2}+1$ & $\frac{S+1}{S+2}$
& $\frac{S+1}{S+2}$ \\
\hline  $\frac{S}{2},\frac{S}{2}-2,...,-\frac{S}{2}+1$ &
$-\frac{S}{2},-\frac{S}{2}+2,...,\frac{S}{2}-1$ & $\frac{S+1}{S+2}$ & $-\frac{S+1}{S+2}$ \\
\hline  $-\frac{S}{2},-\frac{S}{2}+2,...,\frac{S}{2}-1$ &
$\frac{S}{2},\frac{S}{2}-2,...,-\frac{S}{2}+1$ &
$-\frac{S+1}{S+2}$
& $\frac{S+1}{S+2}$ \\
\hline $-\frac{S}{2},-\frac{S}{2}+2,...,\frac{S}{2}-1$ &
$-\frac{S}{2},-\frac{S}{2}+2,...,\frac{S}{2}-1$ & $-\frac{S+1}{S+2}$ & $-\frac{S+1}{S+2}$ \\
\end{tabular}
\end{ruledtabular}}
\end{table}

To consider the ES of the transformed VBS state, we cut the spin chain into
two parts A and B (see Fig.1b). When the edge spin $\tau _{L+1}^{z}$ is
transformed into the $\tau _{L+1}^{x}$ basis, the VBS wave function is
written as {\small 
\begin{eqnarray}
&&|\text{VBS}\left( \alpha ,\beta \right) \rangle
=\sum_{n=0}^{S}\sum_{m=0}^{S}\left( -1\right) ^{S-n}C_{S}^{n}F\left(
S,n,m\right)   \nonumber \\
&&\text{ \ \ \ \ }\times |\text{VBS}\left( \tau _{1}^{z}=\alpha ,\tau
_{L}^{z}=n-S/2\right) \rangle _{A}  \nonumber \\
&&\text{ \ \ \ \ }\times |\text{VBS}\left( \tau _{L+1}^{x}=m-S/2,\tau
_{N}^{x}=\beta \right) \rangle _{B},
\end{eqnarray}%
}where{\small 
\[
F\left( S,n,m\right) =\sum_{p=0}^{S-n}\sum_{q=0}^{n}\frac{(-1)^{n-q}}{2^{S/2}%
}C_{S-n}^{p}C_{n}^{q}\delta _{m,p+q},
\]%
}and we have expressed the edge spins in the wave function for part A in
terms of $\tau _{1}^{z}$ and $\tau _{L}^{z}$ basis, and the edge spins in
the wave function of part B in terms of $\tau _{L+1}^{x}$ and $\tau _{N}^{x}$
basis.

When the non-local unitary transformation is applied to the VBS wave
function, we separate the transformation into three parts $%
U=U_{A}U_{B}U_{AB} $, where {\small 
\[
U_{A}=\prod_{j<k\in A}e^{i\pi s_{j}^{z}s_{k}^{x}},U_{B}=\prod_{j<k\in
B}e^{i\pi s_{j}^{z}s_{k}^{x}},U_{AB}=e^{i\pi s_{A}^{z}s_{B}^{x}}. 
\]%
}$U_{AB}$ plays the essential role in determining the new ES. $s_{A}^{z}$
and $s_{B}^{x}$ take the eigenvalues of the corresponding edge spins. Then
the transformed wave function is written as {\small 
\begin{eqnarray}
&&|\widetilde{\text{VBS}}\left( \alpha ,\beta \right) \rangle  \nonumber \\
&=&\sum_{n=0}^{S}\sum_{m=0}^{S}\left( -1\right) ^{S-n}C_{S}^{n}F\left(
S,n,m\right) e^{i\pi \left( \alpha +n-\frac{S}{2}\right) \left( m+\beta -%
\frac{S}{2}\right) }  \nonumber \\
&&\text{ \ \ }\times |\widetilde{\text{VBS}}\left( \tau _{1}^{z}=\alpha
,\tau _{L}^{z}=n-S/2\right) \rangle _{A}  \nonumber \\
&&\text{ \ \ }\times |\widetilde{\text{VBS}}\left( \tau
_{L+1}^{x}=m-S/2,\tau _{L}^{x}=\beta \right) \rangle _{B},
\end{eqnarray}%
}and the total density matrix of $|\widetilde{\text{VBS}}\left( \alpha
,\beta \right) \rangle $ is thus obtained. After the normalization factors
are properly taken into account, we can trace out the degrees of freedom of
the wave function of the part A, and the reduced density matrix is derived
as {\small 
\begin{eqnarray}
&&\left[ \mathbf{\rho }_{B}\left( \alpha ,\beta \right) \right]
_{m,m^{\prime }}=\sum_{n=0}^{S}\sum_{m=0}^{S}\sum_{m^{\prime }=0}^{S}\frac{1%
}{S+1}\frac{C_{S}^{n}}{\sqrt{C_{S}^{m}C_{S}^{m^{\prime }}}}  \nonumber \\
&&\text{ \ \ \ \ }\times F\left( S,n,m\right) F\left( S,n,m^{\prime }\right)
e^{i\pi \left( \alpha +n-\frac{S}{2}\right) \left( m-m^{\prime }\right) }.
\end{eqnarray}%
}

To complete the summations, some further algebra is performed to rewrite the
function $F\left( S,n,m\right) F\left( S,n,m^{\prime }\right) $ in terms of
the Schwinger bosons. When $\left( m-m^{\prime }\right) $ is even, the phase
factor disappears $e^{i\pi \left( \alpha +n-\frac{S}{2}\right) \left(
m-m^{\prime }\right) }=1$, and the diagonal elements of the reduced density
matrix is obtained {\small 
\begin{equation}
\left[ \mathbf{\rho }_{B}\left( \alpha ,\beta \right) \right] _{m,m^{\prime
}}=\frac{\delta _{m,m^{\prime }}}{S+1}.
\end{equation}%
} However, when $\left( m-m^{\prime }\right) $ is odd, there is a phase
factor $(-1)^{\left( \alpha +n-\frac{S}{2}\right) }$ in the summations, and
non-diagonal elements of the reduced density matrix is given by{\small 
\begin{equation}
\left[ \mathbf{\rho }_{B}\left( \alpha ,\beta \right) \right] _{m,m^{\prime
}}=\frac{(-1)^{\alpha -S/2}}{\left( S+1\right) }\delta _{m+m^{\prime },S}.
\end{equation}%
}For the even-integer spins $S$, all the non-diagonal matrix elements are
zero because $m+m^{\prime }$ can not be even, and then $\mathbf{\rho }_{B}$
is a diagonal matrix with the same element $1/\left( S+1\right) $. This
implies that the ES of the VBS states with even-integer spins remains the 
\emph{same} after the non-local unitary transformation. Thus the degeneracy
in ES of VBS states with even-integer spins is \emph{not} of topological
nature, consistent with the absence of the non-local SOP found previously
and triviality of this phase.\cite{Pollmann-Oshikawa-2012}

However, for the odd-integer spins $S$, the above results show that $\mathbf{%
\rho }_{B}$ has diagonal elements $1/\left( S+1\right) $ and skew diagonal
elements $(-1)^{\alpha -S/2}/\left( S+1\right) $, so it is a X-form matrix.
For any edge configuration, $\mathbf{\rho }_{B}$ has one eigenvalue $%
2/\left( S+1\right) $, and the entanglement energy level is dramatically
changed and given by $\widetilde{\zeta }=\ln \left( S+1\right) -\ln 2$ with $%
(S+1)/2$ fold degeneracy. In the spin-1 case, the transformed VBS wave
function is a ferromagnetic product state, and the entanglement energy level
is non-degenerate\textit{.}\cite{Kunishi} For the $S>1$ odd-integer spins,
however, the transformed VBS wave functions are still entangled states,
which are spin partially polarized ferromagnetic states with a degeneracy $%
\left( S+1\right) /2$ for each class. The ES degeneracy of the transformed
VBS states depends on the precise form of the resultant ferromagnetic
states, and is thus non-topological. The crucial point, however, is that the
non-local unitary transformation with the discrete symmetry $Z_{2}\times
Z_{2}$ has lifted the two-fold topological degeneracy, which is shared by
all of these odd-integer spin VBS states belonging to the \emph{same}
topological phase\cite{Pollmann-Oshikawa-2012}. Meanwhile, the topological
degeneracy and the topologically protecting symmetry can be read off from
the non-local unitary transformation.

\section{$SO(5)$ symmetric VBS state}

In order to put the above results in a large context, we consider VBS states
with larger symmetry groups which represent SPT phases with higher symmetry.
We start with a topologically \emph{nontrivial} VBS state formed in a spin-$%
2 $ chain. Viewing each spin-$2$ as formed by two spin-$3/2$'s, we can
construct an $SO(5)$ symmetric VBS state\cite{Tu-2008,TZX}. As pointed out
in Ref.\cite{congjun,Tu-2006}, the $\pm 3/2,\pm 1/2$ states of a spin-$3/2$
can be regarded as the four states of the spinor representation of $SO(5)$.
Similarly one can also represent the $\pm 2,\pm 1,0$ states of spin-$2$ as
the five-dimensional vector irreducible representation (IR) of $SO(5)$. Then
we can view the vector IR as the symmetric component of the tensor product
of two virtual spinor IR's, i.e., {\small 
\begin{equation}
\underline{4}\otimes \underline{4}=\underline{1}\oplus \underline{5}\oplus 
\underline{10}.  \label{decom1}
\end{equation}%
}The numerals are the dimensions of the $SO(5)$ IR's. The tensor product of
two $\underline{5}$'s on adjacent sites decomposes into {\small 
\begin{equation}
\underline{5}\otimes \underline{5}=\underline{1}\oplus \underline{10}\oplus 
\underline{14}.  \label{decom2}
\end{equation}%
}Comparing Eqs. (\ref{decom1}) and (\ref{decom2}), we can regard Eq.~(\ref%
{decom1}) as the tensor product of two neighboring virtual spins after their
respective partners have form $SO(5)$ singlet with other virtual spins. Then
one can find that $SO(5)$ singlet $\underline{1}$ and the antisymmetric $%
\underline{10}$ appear in the decomposition but the symmetric $\underline{14}
$ is absent. Therefore, if $H=\sum_{i}P_{14}(i,i+1)$, the $SO(5)$ symmetric
VBS state where neighboring virtual $\underline{4}$'s pair into $SO(5)$
singlet will be the ground state. The operator $P_{\underline{14}}(i,i+1)$
can be expressed in terms of the $SO(5)$ generators {\small 
\begin{equation}
P_{\underline{14}}(i,j)=\frac{1}{2}\sum_{1\leq a<b\leq
5}L_{i}^{ab}L_{j}^{ab}+\frac{1}{10}(\sum_{1\leq a<b\leq
b}L_{i}^{ab}L_{j}^{ab})^{2}+\frac{1}{5}.
\end{equation}%
}Because the physical spin $SU(2)$ is a subgroup of $SO(5)$, each IR of $%
SO(5)$ must decompose into an integral number of $SU(2)$ multiplets. Thus
the $\underline{14}$ discussed above must be expressible as the direct sum
of $SU(2)$ IR obtained by decomposing the direct product of two $S=2$
multiplets. Since the 14-dimensional IR is symmetric upon the exchange of
site indices, it must only contain even-spin $SU(2)$ multiplets, i.e. $%
\underline{14}\rightarrow S_{t}=2\oplus S_{t}=4$.

For an open chain, there exist a free spin-$3/2$ at each end of the chain,
leading to 4-fold degeneracy which is topologically protected by a larger
symmetry. The corresponding topological ES degeneracy should also be 4-fold.
For comparison, in the $SO(3)$ symmetric VBS states, topology only protects
a half edge spin and 2-fold ES degeneracy for odd-integer spins, while there
is no protection at all for even-integer spins.

\section{$SO(2S+1)$ symmetric VBS states with integer spins}

It is straightforward to promote the symmetry of the VBS states and demand
that the spin-$S$ states on each site transform under the $\left(
2S+1\right) $-dimensional vector representation of $SO\left( 2S+1\right) $,
which can be formed by tensor decomposition of two virtual $2^{S}$%
-dimensional spinors.\cite{Tu-2008,TZX,Tu-2009} The main issue here is to
identify the topological degeneracy. Since $SO(2S+1)$ is a rank-$S$ algebra,
there are $S$ mutually commuting Cartan generators: $\{L^{12},L^{34},\ldots
,L^{2S-1,2S}\}$. At each site, the quantum states are classified by the
eigenvalues of these Cartan generators as {\small 
\begin{equation}
L^{2\alpha -1,2\alpha }|m_{\alpha }\rangle =m_{\alpha }|m_{\alpha }\rangle
,\quad (m_{\alpha }=0,\pm 1).
\end{equation}%
}Thus the single-site states are associated with $S$ quantum numbers $%
\{m_{1},\cdots ,m_{S}\}$, and they are subjected to the constraint {\small 
\begin{equation}
m_{\alpha }m_{\beta }=0,\qquad (\alpha \not=\beta ).  \label{eq:constraint}
\end{equation}%
}The topological feature of these VBS states can be characterized by the
following generalized non-local SOP {\small 
\begin{equation}
\mathcal{O}^{ab}=\lim_{|j-i|\rightarrow \infty }\langle
L_{i}^{ab}\prod_{r=i}^{j-1}\exp (i\pi L_{r}^{ab})L_{j}^{ab}\rangle .
\label{eq:SOP}
\end{equation}%
}Since the ground state is $SO(2S+1)$ rotational invariant, the above
non-local order parameters should all be equal to each other. To determine
the value of these parameters, only $\mathcal{O}^{12}$ needs to be
evaluated. In the $L^{12}$ channel, the role of the phase factor in Eq. (\ref%
{eq:SOP}) is to correlate the finite spin polarized states in the $m_{1}$
channel at the two ends of the string. If nonzero $m_{1}$ takes the same
value at the two ends, then the phase factor is equal to $1$. On the other
hand, if nonzero $m_{1}$ takes two different values at the two ends, then
the phase factor is equal to $-1$. Thus the value of $\mathcal{O}^{12}$ is
determined purely by the probability of $m_{1}=\pm 1$ appearing at the two
ends of the string. It is straightforward to show that the probability of
the states $m_{1}=\pm 1$ appearing at one lattice site is $2/(2S+1)$ and
thus $\mathcal{O}^{12}=4/(2S+1)^{2}$.

In the $SO(2S+1)$ Lie algebra, $(L^{2\alpha -1,2\alpha },L^{2\alpha
-1,2S+1}, $ $L^{2\alpha ,2S+1})$ span an $SO(3)$ sub-algebra in which $\exp
(i\pi L^{2\alpha ,2S+1})$ plays the role of flipping the quantum number $%
m_{\alpha }$. This exponential operator can flip the quantum numbers of $%
m_{\alpha }$ without disturbing the quantum states in all other channels.
Thus the following non-local unitary transformation with $Z_{2}\times Z_{2}$
discrete symmetry {\small 
\begin{equation}
U_{\alpha }=\prod_{j<i}\exp \left( i\pi L_{j}^{2\alpha -1,2\alpha
}L_{i}^{2\alpha ,2S+1}\right) ,
\end{equation}%
}can change the spin configuration in the $m_{\alpha }$ channel into a
ferromagnetically ordered one. Furthermore, by performing this non-local
transformation successively in all the channels $U=\prod_{\alpha
=1}^{S}U_{\alpha }$, all the configurations of the ground state will become
ferromagnetically ordered. Applying $U$ to the \textit{Cartan} generators,
it can be shown that {\small 
\begin{equation}
UL_{i}^{ab}U^{-1}=L_{i}^{ab}\exp (i\pi \sum_{j=1}^{i-1}L_{j}^{ab}).
\end{equation}%
}Substituting this formula to Eq. (\ref{eq:SOP}), we find that {\small 
\begin{equation}
\mathcal{O}^{ab}=\lim_{\left| j-i\right| \rightarrow \infty }\left\langle
L_{i}^{ab}L_{j}^{ab}\right\rangle _{U}.  \label{eq:CF}
\end{equation}%
}Thus the non-local SOP $\mathcal{O}^{ab}$ of the Cartan generators become
the ordinary two-point correlation functions of local operators. Under this
general non-local unitary transformation with the $\left( Z_{2}\times
Z_{2}\right) ^{S}$ symmetry, the topological order of the original VBS
states is transformed into conventional ferromagnetic order, just like in
the SO(3) case.

When an open chain system is considered, there appear $2^{S}$ degenerate
edge states at each end of the chain, which are also topologically protected
by the discrete symmetry $(Z_{2}\times Z_{2})^{S}$. Accordingly the
topological degeneracy in the ES can be read off as $2^{S}$-fold. We thus
find that a higher symmetry in the SPT phase protects a higher ES degeneracy.

\section{Conclusion}

We have shown through explicit examples that entanglement spectra of
topological ground states contain both universal and non-universal
structures. The topological degeneracy of the lowest entanglement energy
level may be lifted or isolated by a generalized non-local unitary
transformation or topological disentangler, which is determined by the
minimal symmetry protecting the topological phases. Our results shed
significant light on the issue of using entanglement spectrum to identify
topological order.

\section{Acknowledgements}

One of the authors (GMZ) would like to thank H. H. Tu and T. Xiang for their
earlier collaborations and acknowledge the support of NSF-China through the
grant No.20121302227. KY is supported by National Science Foundation through
grant No. DMR-1004545.


\begin{thebibliography}{99}
\bibitem{Kitaev-Preskill} A. Kitaev and J. Preskill, Phys. Rev. Lett. 
\textbf{96}, 110404 (2006).

\bibitem{Levin-Wen} M. Levin and X. G. Wen, Phys. Rev. Lett. \textbf{96},
110405 (2006).

\bibitem{Zhang-Vishwanath} Y. Zhang, T. Grover, A. Turner, M. Oshikawa, and
A. Vishwanath, Phys. Rev. B \textbf{85}, 235151 (2012).

\bibitem{Li-Haldane} H. Li and F. D. M. Haldane, Phys. Rev. Lett. \textbf{101%
}, 010504 (2008).

\bibitem{Fidkowski} L. Fidkowski, Phys. Rev. Lett. \textbf{104}, 130502
(2010).

\bibitem{cirac-2011} J. I. Cirac, D. Poiblanc, N. Schuch, and F. Verstraete,
Phys. Rev. B \textbf{83}, 245134 (2011).

\bibitem{Bernevig-2011} A. Chandran, M. Hermanns, N. Regnault, and B. A.
Bernevig, Phys. Rev. B \textbf{84}, 205136 (2011).

\bibitem{Qi-2012} X. L. Qi, H. Katsura, and A. W. W. Ludwig, Phys. Rev.
Lett. \textbf{108}, 196402 (2012).

\bibitem{Chandran-Sondhi} A. Chandran, V. Khemani, and S. L. Sondhi,
arXiv:1311.2946.

\bibitem{Chen-Gu-Wen-2011} X. Chen, Z. C. Gu, and X. G. Wen, Phys. Rev. B 
\textbf{83}, 035107 (2011).

\bibitem{Schuch} N. Schuch, D. Perez-Garcia, and I. Cirac, Phys. Rev. B 
\textbf{84}, 165139 (2011).

\bibitem{chen-gu-liu-wen} X. Chen, Z. C. Gu, Z. X. Liu, and X. G. Wen, Phys.
Rev. B \textbf{87}, 155114 (2013).

\bibitem{Quella} K. Duivenvoorden and T. Quella, Phys. Rev. B \textbf{87},
125145 (2013).

\bibitem{Katsura-2007} H. Katsura, T. Hirano, and Y. Hatsugai, Phys. Rev. B 
\textbf{76}, 012401 (2007).

\bibitem{Haldane-1983} F. D. M. Haldane, Phys. Lett. \textbf{93A}, 464
(1983); Phys. Rev. Lett. \textbf{50}, 1153 (1983).

\bibitem{Gu-Wen-2009} Z. C. Gu and X. G. Wen, Phys. Rev. B \textbf{80},
155131 (2009).

\bibitem{Pollmann-Oshikawa-2010} F. Pollmann, A. M. Turner, E. Berg, and M.
Oshikawa, Phys. Rev. B \textbf{81}, 064439 (2010).

\bibitem{Pollmann-Oshikawa-2012} F. Pollmann, E. Berg, A. M. Turner, and M.
Oshikawa, Phys. Rev. B \textbf{85}, 075125 (2012).

\bibitem{AKLT} I. Affleck, T. Kennedy, E. H. Lieb, and H. Tasaki, Phys. Rev.
Lett. \textbf{59}, 799 (1987); Commun. Math. Phys. \textbf{115}, 477 (1988).

\bibitem{Tu-2008} H. H. Tu, G. M. Zhang, and T. Xiang, J. Phys. A: Math.
Theor. \textbf{41}, 415201 (2008).

\bibitem{TZX} H. H. Tu, G. M. Zhang, and T. Xiang, Phys. Rev. B \textbf{78},
094404 (2008).

\bibitem{Tu-2009} H. H. Tu, G. M. Zhang, T. Xiang, Z. X. Liu, and T. K. Ng,
Phys. Rev. B \textbf{80}, 014401 (2009).

\bibitem{Else-Bartlett-Doherty} D. V. Else, S. D. Bartlett, and A. C.
Doherty, Phys. Rev. B \textbf{88}, 085114 (2013).

\bibitem{Quella-2013} K. Duivenvoorden and T. Quella, Phys. Rev. B \textbf{88%
}, 125115 (2013).

\bibitem{Arovas-Haldane-1988} D. P. Arovas, A. Auerbach, and F. D. M.
Haldane, Phys. Rev. Lett. \textbf{60}, 531 (1988).

\bibitem{Nijs-Rommelse} M. den Nijs and K. Rommelse, Phys. Rev. B \textbf{40}%
, 4709 (1989).

\bibitem{Oshikawa-1992} M. Oshikawa, J. Phys.: Condens. Matter \textbf{4},
7469 (1992).

\bibitem{Totsuka-Suzuki} K. Totsuka and M. Suzuki, J. Phys.:Condens Matter 
\textbf{7}, 1639 (1995).

\bibitem{Kennedy-Tasaki-1992} T. Kennedy and H. Tasaki, Phys. Rev. B \textbf{%
45}, 304 (1992); Commun. Math. Phys. \textbf{147}, 431 (1992).

\bibitem{Rao-Zhang-Yang} W. J. Rao, G. M. Zhang, and K. Yang, in preparation.

\bibitem{Kunishi} K. Okunishi, Phys. Rev. B \textbf{83}, 104411 (2011).

\bibitem{congjun} C. J. Wu, J. P. Hu, and S. C. Zhang, Phys. Rev. Lett. 
\textbf{91}, 186402 (2003).

\bibitem{Tu-2006} H. H. Tu, G. M. Zhang, and L. Yu, Phys. Rev. B \textbf{74}%
, 174404 (2006); \textbf{76},014438 (2007).
\end{thebibliography}
\end{document}